\documentclass[sigconf]{acmart}

\usepackage{longtable}
\usepackage{multirow}
\usepackage{color}
\usepackage{float}
\usepackage{longtable}
\usepackage{stfloats}
\usepackage{enumitem}
\usepackage{blindtext}
\usepackage{etoolbox}
\usepackage{graphicx}
\usepackage{subfigure}

\AtBeginDocument{%
  \providecommand\BibTeX{{%
    \normalfont B\kern-0.5em{\scshape i\kern-0.25em b}\kern-0.8em\TeX}}}

\copyrightyear{2024}
\acmYear{2024}
\setcopyright{acmlicensed}\acmConference[KDD '24]{Proceedings of the 30th ACM SIGKDD Conference on Knowledge Discovery and Data Mining}{August 25--29, 2024}{Barcelona, Spain}
\acmBooktitle{Proceedings of the 30th ACM SIGKDD Conference on Knowledge Discovery and Data Mining (KDD '24), August 25--29, 2024, Barcelona, Spain}
\acmDOI{10.1145/3637528.3671621}
\acmISBN{979-8-4007-0490-1/24/08}
\acmPrice{}

\begin{document}


\title{COMET: NFT Price Prediction with Wallet Profiling}



\author{Tianfu Wang}
\affiliation{%
  \institution{School of Computer Science and Technology, University of Science and Technology of China}
  \city{Hefei}
  \country{China}
}
\email{tianfuwang@mail.ustc.edu.cn}

\author{Liwei Deng}
\affiliation{%
  \institution{Microsoft Inc.}
  \city{Suzhou}
  \country{China}
}
\email{denglw0830@outlook.com}

\author{Chao Wang}
\affiliation{%
  \institution{School of Artificial Intelligence and Data Science, University of Science and Technology of China}
  \city{Hefei}
  \country{China}
}
\email{chadwang2012@gmail.com}



\author{Jianxun Lian}
\affiliation{%
  \institution{Microsoft Research Asia}
  \city{Beijing}
  \country{China}
}
\email{jianxun.lian@outlook.com}

\author{Yue Yan}
\affiliation{
 \institution{Microsoft Inc.}
  \city{Beijing}
  \country{China}
}
\email{yue.yan@microsoft.com}

\author{Nicholas Jing Yuan}
\authornote{Corresponding authors.}
\affiliation{%
  \institution{Microsoft Inc.}
  \city{Suzhou}
  \country{China}
  }
\email{nicholas.yuan@microsoft.com}

\author{Qi Zhang}
\affiliation{%
  \institution{Microsoft Inc.}
  \city{Suzhou}
  \country{China}
}
\email{zhang.qi@microsoft.com}

\author{Hui Xiong}
\authornotemark[1]
\affiliation{%
  \institution{Thrust of Artificial Intelligence, The Hong Kong University of Science and Technology (Guangzhou)}
  \city{Guangzhou}
  \country{China}
}
\affiliation{%
  \institution{Department of Computer Science and Engineering, The Hong Kong University of Science and Technology}
  \city{Hong Kong SAR}
  \country{China}
 }
\email{xionghui@ust.hk}

\renewcommand{\shortauthors}{Tianfu Wang et al.}

\begin{abstract}
As the non-fungible token (NFT) market flourishes, price prediction emerges as a pivotal direction for investors gaining valuable insight to maximize returns. However, existing works suffer from a lack of practical definitions and standardized evaluations, limiting their practical application. Moreover, the influence of users’ multi-behaviour transactions that are publicly accessible on NFT price is still not explored and exhibits challenges. In this paper, we address these gaps by presenting a practical and hierarchical problem definition. This approach unifies both collection-level and token-level task and evaluation methods, which cater to varied practical requirements of investors. To further understand the impact of user behaviours on the variation of NFT price, we propose a general wallet profiling framework and develop a \textbf{CO}mmunity enhanced \textbf{M}ulti-b\textbf{E}havior \textbf{T}ransaction graph model, named \textbf{COMET}. COMET profiles wallets with a comprehensive view and considers the impact of diverse relations and interactions within the NFT ecosystem on NFT price variations, thereby improving prediction performance. Extensive experiments conducted in our deployed system demonstrate the superiority of COMET, underscoring its potential in the insight toolkit for NFT investors.
\end{abstract}

\ccsdesc[500]{Information systems~Data mining}

\keywords{Web3, Non-fungible Token, FinTech, Graph Learning}


\received{20 February 2007}
\received[revised]{12 March 2009}
\received[accepted]{5 June 2009}

\maketitle

\section{Introduction}
\begin{figure}[t]
    \centering
    \includegraphics[width=0.46\textwidth]{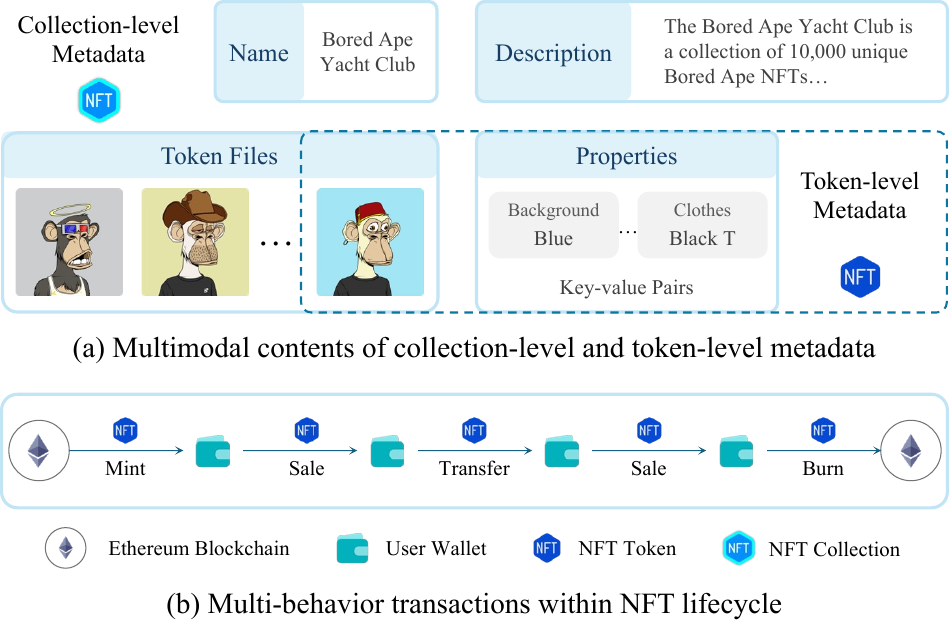}
    \vspace{-0.15cm}
    \caption{An illustration of NFT ecosystem. (a) Multimodal contents: collection-level name and description, and token-level file and properties. (b) Multi-behaviour transactions: mint, sale, transfer, and burn.}
    \label{fig:nft-ecosystem}
    \vspace{-0.15cm}
\end{figure}

In recent years, the advent of blockchain technology and Web3 has given rise to a novel class of digital assets,  known as non-fungible tokens (NFTs)\cite{web3-www-2023-evolution}.
Unlike cryptocurrencies, such as Bitcoin and ETH, NFTs are media files (e.g., visual content and text descriptions as shown in Figure~\ref{fig:nft-ecosystem}(a)), attached to a specific digital token that can be tracked and verified on a blockchain, which endows the NFTs the property of investment as digital collection. 
According to Forbes\footnote{https://www.forbes.com/digital-assets/nft-prices} statistics, as of February 2024, the global NFT market cap has reached nearly 18.29 billion dollars.
It signifies a remarkable growth rate of about 2,636\% compared to the figures in February 2021.
The fastly developing market brings a lot of profit opportunities, which attracts many investors into the wave of investment. 
Therefore, estimating NFT price, offering valuable insights to guide the investment process, draws lots of effort. 
Existing studies for estimating NFT price can be roughly divided into two directions, i.e., NFT asset valuation and NFT price prediction, based on different ideas to get benefits. 
NFT asset valuation refers to discovering the real potential value of NFT, aiming to look for underestimated high-value NFT.
Existing approaches mine various factors that can influence the worth of NFTs. For example, Mekacher et al.~\cite{valuation-naturesp-2022-heterogeneous} show that tokens with rare or distinctive properties can create a sense of scarcity, often resulting in higher prices. 
Costa et al.~\cite{valuation-www-2023-multimodal} predict the NTF price based on its visual content.  
However, these works fail to consider the dynamic nature of NFT market, making it difficult to suit the market changing with each passing day. As a result, To better adapt to the highly dynamic market environment, more researchers are focusing on NFT price prediction~\cite{valuation-goodit-2022-pictures,valuation-goodit-2023-keword,valuation-arxiv-2022-appraisal}. It aims to forecast the daily prices of NFTs and pay attention to the rise and fall of prices over time, 
studies in this direction typically adopt a sequential model to extract useful information from the historical sequence of prices. For example, Jain et al.~\cite{valuation-arxiv-2022-appraisal} leverage recurrent neural networks (RNNs) to forecast the daily price of a collection based on past prices. 
Luo et al.~\cite{valuation-goodit-2023-keword} incorporate twitter activities as supplementary features to predict future prices of single collections.

Despite growing interest in NFTs, existing studies for price prediction are still limited to the following aspects. 
On the one hand, existing works are characterized by ambiguous and incomplete problem definitions and evaluation methods, making it difficult to fairly compare approaches and effectively advance the field. 
We organize the recent studies in Table~\ref{tab:comparison}, which highlights their discrepancies in multiple aspects ranging from problem definition to evaluation.
We observe that these works on price prediction only focus on future collection-level prices, while the price variations among tokens within identical collections pose an intriguing challenge that needs further exploration. 
On the other hand, current research on NFT price prediction has failed to leverage valuable information from publicly accessible user behaviors, which offers insights into uncovering potential price trends.
As shown in Figure~\ref{fig:nft-ecosystem} (b), web3 users engage with NFTs through their digital wallets, participating in multi-behavior transactions such as mints (primary trade), sales (secondary trade), transfers (zero-price exchange), and even burns (token destruction). 
These wallet behaviors contain diverse semantics on market sentiments, which have an important impact on shaping market dynamics and NFT prices.
Furthermore, identifying potential price trend correlations among NFTs and between NFTs and wallets also is a significant task.
This allows us to gain deeper insights into social influences and temporal correlations, contributing to enhancing prediction accuracy. For instance, if a wallet with a history of successful NFT investments suddenly starts divesting from a particular collection, it might signal a shift in market sentiment that can reflect price trends.

To tackle these problems, we provide a hierarchical definition of the NFT price prediction problem that is more practical to provide clarity.
Concretely, we formulate the NFT price prediction with forecasting awareness at both the collection and token levels, which enables our framework to cater to the diverse needs of practical investors in such a highly dynamic market. 
To understand the effect of users' multi-behaviour transactions on the variation of NFT price, we propose a general wallet profiling framework that aims to incorporate the influence of wallets when making predictions. Considering the complexity (e.g., multi-behaviour patterns) and sparsity of transactions (e.g., a limited number of wallets participating in trades actively), we develop a \textbf{CO}munity enhanced \textbf{M}uti-b\textbf{E}havior \textbf{T}ransaction graph network model under this framework, named \textbf{COMET}. Specifically, we develop a temporal transaction graph construction method, in which wallets and NFT collections are modeled as nodes and the multi-behaviour transactions are encoded as heterogeneous edges in the constructed graph. It facilitates a comprehensive grasp of the interactions between collections and wallets, simultaneously capturing underlying correlations among different collections. Furthermore, we design a temporal transaction graph learning model to learn the representations of collections and wallets, which can adaptively propagate the influence of wallets on the constructed graph. To further relieve the sparsity problem, we introduce communities to enrich the similarity among the nodes that exist in the same community. Finally, we optimize the proposed model hierarchically for the collection and token level due to the discrepancy, which cannot be ignored, between these two tasks. 

In summary, our contributions are as follows:
\begin{itemize}
    \item We systematically formalize a comprehensive and practical problem definition for NFT price prediction, incorporating both collection-level and token-level analyses, to address the universal requirements of investors.
    \item We propose a novel wallet profiling framework for NFT price prediction, mining the social influence behind transactions.
    Specifically, we develop a graph modeling method, design a community enhanced multi-behaviour transaction graph model, and hierarchically predict two-level targets.
    \item We build an NFT price prediction system and deploy it in practice, which encompasses data infrastructure and product applications. We conduct extensive experiments to validate the effectiveness of COMET and its components.
\end{itemize}

\section{Related Work}
We survey the related studies on NFT valuation and price prediction, graph learning for finance, and temporal graph learning. 

\begin{table*}[t]
\small
\centering
\caption{Comparison of existing works and COMET.}
\begin{tabular}{c|ccc|c|cl}
\toprule
    \multicolumn{1}{c|}{\multirow{2}{*}{Work}} & \multicolumn{3}{c|}{Problem Definition} & \multicolumn{1}{c|}{Methodology} & \multicolumn{2}{c}{Evaluation}  \\ \cline{2-7} 
    \multicolumn{1}{c|}{}  & \multicolumn{1}{c}{Price-level} & \multicolumn{1}{c}{Target} & \multicolumn{1}{c|}{Task Type} & \multicolumn{1}{c|}{} & \multicolumn{1}{c}{Dataset} & \multicolumn{1}{c}{Metrics}  \\
\midrule
\cite{valuation-naturesp-2022-heterogeneous} & Token & Asset Value &  Correlation & Correlation Analysis & 1.4M Tokens & Fitting error \\
\cite{graph-naturesp-2022-crypto-networt} & Token & Asset Value & Correlation & Correlation Analysis  & 48K Tokens & \(R^2\) \\
\cite{valuation-www-2023-multimodal} & Token & Asset Value &  Classification & Multimodal Learning & 4.7M Tokens & Precision; Recall; F1 \\ 
\cite{valuation-www-2023-tweetboost} & Token & Asset Value &  Classification & LightGBM; XGBoost; etc & 62K Tokens & ACC; F1 \\ 
\cite{overview-naturesp-2021-revolution} & Token & Asset Value &  Regression & Linear Regression  & 4.7M Tokens & $R^2$ \\ 
\cite{valuation-goodit-2022-pictures} & Collection & Daily Price &  Correlation & Correlation Analysis & 4 Collections & / \\ 
\cite{valuation-goodit-2023-keword} & Collection & Daily Price &  Regression & NLP sentiment and MLP & 19 Collections & MAE; ACC; F1 \\ 
\cite{valuation-arxiv-2022-appraisal} & Collection & Daily Price &  Regression & RNN; Linear regression & 1 Collections & MSE; ACC \\ 
\midrule
\multirow{2}{*}{\textbf{Ours}} & Collection & Aggregated Price &  Classification & Temporal Heterogeneous & 100 Collections & ACC; MCC \\ 
\multicolumn{1}{c|}{} & Token & Sale Price &  Regression & Graph Learning & 1.5M Tokens & MSE; MAE  \\ 
\bottomrule
\end{tabular}
\label{tab:comparison}
\end{table*}

\textbf{NFT Valuation and Price Prediction.} 
As an emerging digital asset class, NFTs have attracted exploratory efforts that consider various influencing factors aimed at estimating their prices.
We categorize them into asset valuation and price prediction, summarized in Table \ref{tab:comparison}.
From this table, we can observe that the asset value approaches are all for token-level NFT pricing, while the price prediction methods are for daily-level collection price prediction. The former usually estimates the current or intrinsic value of a specific NFT based on multiple relatively static factors, such as rarity~\cite{valuation-naturesp-2022-heterogeneous}, multimodal content~\cite{valuation-www-2023-multimodal}, and twitter activities~\cite{valuation-www-2023-tweetboost}. 
However, these works lack forecasting awareness and cannot meet the requirements of investors in such a dynamic market.
Existing studies on NFT price prediction mainly focus on daily collection price prediction. For example, Brunet et al.~\cite{valuation-goodit-2022-pictures} collected tweets based on keyword search and analyzed the latent correlation between the number of tweets and the average collection price. Luo et al.~\cite{valuation-goodit-2023-keword}
conducted sentiment analysis based on tweet texts and predicted the next-day movement of the collection-level price with multiple regression models. 
However, due to the lack of standardized problem definitions, datasets and evaluations, these approaches are hard to be fairly compared with each other. Moreover, they can only provide vague signals on which collection should be invested, while ignoring the discrepancy between a specific token and collections, e.g., the price trend of an NFT may be contrary to that of the whole collection. In addition, the influence of wallets' multi-behaviours on the collection price is not included when modeling the predictive patterns. These shortcomings prevent the existing approaches from achieving the optima. Different from previous works, we present a practical
problem definition of NFT price prediction and a wallet profiling framework to analyze the social influence on NFT prices.

\textbf{Graph Learning for Finance.} Graph structure plays an important role in presenting financial knowledge, which facilitates understanding the correlations among financial assets~\cite{stcok-arxiv-2023-survey}. 
A notable application of this can be found in stock price prediction. Works \cite{stock-ijcai-2020-graph,stock-www-2021-rest} constructed stock relationship graphs based on historical sequence correlations or stock description documents. Furthermore, studies \cite{stcok-kdd-2019-indicator,stcok-ksme-2023-hybrid} explored the relationship between stocks and fund managers by constructing a heterogeneous graph. Utilizing the constructed graphs, most of these studies employ graph neural networks (GNNs) to extract meaningful features for their financial tasks~\cite{chenliyi2024collaboration}.
Despite the prevalence of graph structures in stock markets, the construction and modeling methods for graphs remain unexplored in NFT price prediction. Incorporating multi-behavior transactions into graphs is a challenge that needs to be addressed.

\textbf{Temporal Graph Learning. }
As graph learning has demonstrated impressive performance on static graphs~\cite{zhengzhi-2021drug,tianfuwang-ijcai-2024-flag-vne,zhengzhi-2023interaction,dingleilei-ijcai-2024-gnn-dgr}, researchers have started to explore more challenging graphs with temporal evolution. 
The paradigms of existing methods can be broadly grouped into timestamp embedding, parameter sequential embedding, and hidden state sequential embedding~\cite{dl-arxiv-2023-dynamic-gnn-survey}. Timestamp embedding-base methods~\cite{gnn-www-2020-tdgnn} treat timestamps as a feature type, incorporating them alongside other features during the propagation process of GNNs. 
Methods based on parameter sequential embedding~\cite{gnn-aaai-2020-egcn} use RNNs to dynamically encode the parameters of GNNs across time steps.
Approaches based on hidden state sequential embedding~\cite{gnn-kdd-2019-path} encode each snapshot of data with GNNs, then employ RNNs to capture the temporal relationships between these snapshots.
Although these methods are adept at capturing temporal information, they struggle with addressing challenges posed by sparse and multi-behaviour transactions in NFT domains.

\section{Problem Definition}
\textbf{Notations.} 
Within in NFT ecosystem, let us denote a set of NFT collections as $\mathcal{C}$. For each collection $c \in \mathcal{C}$, we denote a set of its tokens as $\mathcal{U}_{c}$. 
There is also a set of wallets, and we denote it as $\mathcal{W}$. For a token $u \in \mathcal{U}_c$ 
, there may be a sequence of historical sale transactions, we denote it as $\mathcal{S}_u$.

To bring clarity to this evolving field, we present a more practical and hierarchical definition of the NFT price prediction problem. Following the guide for the finance application \cite{finance-book-1999-market}, our definition covers both the collection-level and the token-level tasks. This approach not only allows us to analyze the overall price trends of collections, but also understand the relative price relationships between individual tokens and their collections.

\textbf{Collection-Level price trend prediction.} 
Given the high-frequency natures of collection-level prices, predicting its trend is particularly instructive to investors \cite{stcok-arxiv-2023-survey}, which allows investors to stay informed about the changing trends.
Similar to studies on stock movement predictions \cite{stock-www-2021-rest,stock-kdd-2021-accurate},
we focus on predicting the collection-level price trend over a specified period. 
At time step $T$, we consider historical data within a window of size $H$ denoted as $\{\mathcal{X}_c^{t} \mid t \in [T-H, T]\}$, in which $\mathcal{X}_c^{t}$ is not constrained to be the historical price and can be various factors related to the NFT prices (e.g., daily transaction counts and total sale volume). 
To predict the price movement after the prediction step $N$, we define the binary variable $y_c$ as follows: $y_c = 1$ if $p_c^{T + N} - p_c^{T} > 0$, otherwise $y_c = 0$. Here, $p_c^t$ represents the collection-level price at time step $t$.
Our objective is to build a model that predicts the future movement of collection-level price:
\begin{equation}
    \hat{y}_c = \mathcal{F}_c(\{\mathcal{X}^{t}_c \mid t \in [T-H, T]\}).
\end{equation}

\textbf{Token-Level sale price prediction.} Sales of individual NFT tokens are typically infrequent and may have long intervals between them.
Drawing inspiration from works on real estate \cite{real-kdd-16-days,real-kdd-2021-mugrep}, we aim to estimate future sale prices at which individual NFT tokens are sold on the market, denoted as $y_u = p_u^{T + N}$. In addition to collection-level historical data that contains the overall trend of the entire collection, we incorporate the historical sale transactions $S_u$ of each token as input to consider the token's relative relationship with the overall collections. This predictive model is defined as:
\begin{equation}
    \hat{y}_u = \mathcal{F}_u\left(S_u; \mathcal{F}_c\left(\{\mathcal{X}^{t}_c \mid t \in [T-H, T]\}\right)\right).
\end{equation}

Overall, our proposed method unifies both existing token-level works \cite{valuation-naturesp-2022-heterogeneous,graph-naturesp-2022-crypto-networt,valuation-www-2023-multimodal,valuation-www-2023-tweetboost,overview-naturesp-2021-revolution} and collection-level studies \cite{valuation-goodit-2022-pictures,valuation-goodit-2023-keword,valuation-arxiv-2022-appraisal}, while also enabling them to consider future predictability.
The collection-level task provides a broader perspective on overall trends from a macroeconomic perspective, while the token-level task offers detailed insights into specific decisions.
This dual-task approach effectively caters to the diverse needs of investors in the NFT market, enhancing their decision-making capabilities.

\begin{figure*}
    \centering
    \includegraphics[width=1.0\textwidth]{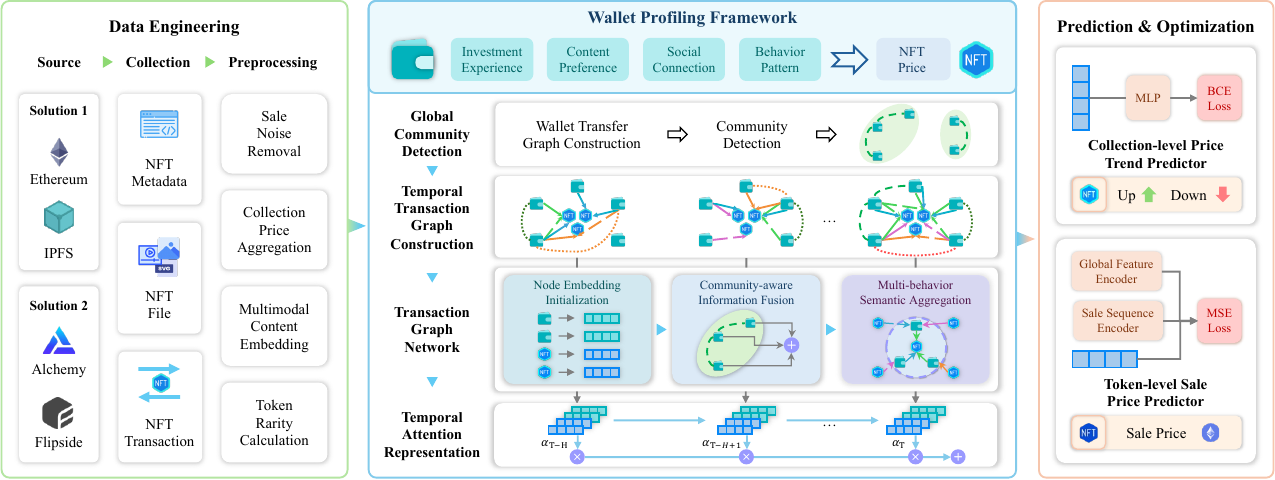}
    \caption{An overview of NFT price prediction system with wallet profiling.}
    \label{fig:enter-label}
\end{figure*}

\section{NFT Price Prediction System}
We build an NFT price prediction system, as shown in Figure~\ref{fig:enter-label}, which consists of three components, i.e., data engineering, model architecture, and prediction \& optimization. Data engineering is used to acquire and preprocess NFT-related data, which provides a data foundation for the following analysis components. Then, the model architecture accepts the processed data features as input and aims to capture the price trend patterns under the wallet profiling framework. Specifically, we instantiate the framework and propose a \textbf{CO}mmunity-enhanced \textbf{M}ulti-b\textbf{E}havior \textbf{T}ransaction graph model, named \textbf{COMET}, which considers the multimodal content and multi-behaviour transactions within the NFT ecosystem. Finally, the prediction \& optimization is adopted to train the model parameters and infer the price in the future. 

\subsection{Data Engineering}
\label{data-engineering}
We collect NFT-related data from multiple resources and perform several key pre-processing to obtain insightful data.

\subsubsection{\text{Data Collection}} 
NFT-related data can be categorized into on-chain and off-chain data. Generally, on-chain data include transaction history, token ownership, etc., which are directly recorded on the blockchain.
In contrast, off-chain data refer to information that exists outside the blockchain, mainly including NFT metadata, NFT files, etc.
We present two distinct data collection solutions that are tailored for both industrial and lighting applications.

\textit{Infrasturture-supported Solution}. We establish a local Ethereum node using Geth\footnote{https://geth.ethereum.org/}, enabling us to extract transaction data directly from the blockchain using the Ethereum ETL\footnote{https://github.com/blockchain-etl/ethereum-etl} library. For off-chain data, we harness the InterPlanetary File System (IPFS)\footnote{https://ipfs.tech/} to crawl them. This industrial-scale solution prioritizes scalability, efficiency, and data integrity. It ensures efficient access to both on-chain and off-chain NFT data, while enabling real-time and unlimited data extraction and handling.

\textit{Public API-based Solution}. We also create an open-source alternative based on publicly accessible API. Concretely, we use Alchemy\footnote{https://www.alchemy.com} and Flipside\footnote{https://flipsidecrypto.xyz/} to obtain both after-processed on-chain and off-chain data. This highly flexible solution provides users and communities with accessible tools for investigating studies related to NFT.

\begin{figure}
    \centering
    \includegraphics[width=0.48\textwidth]{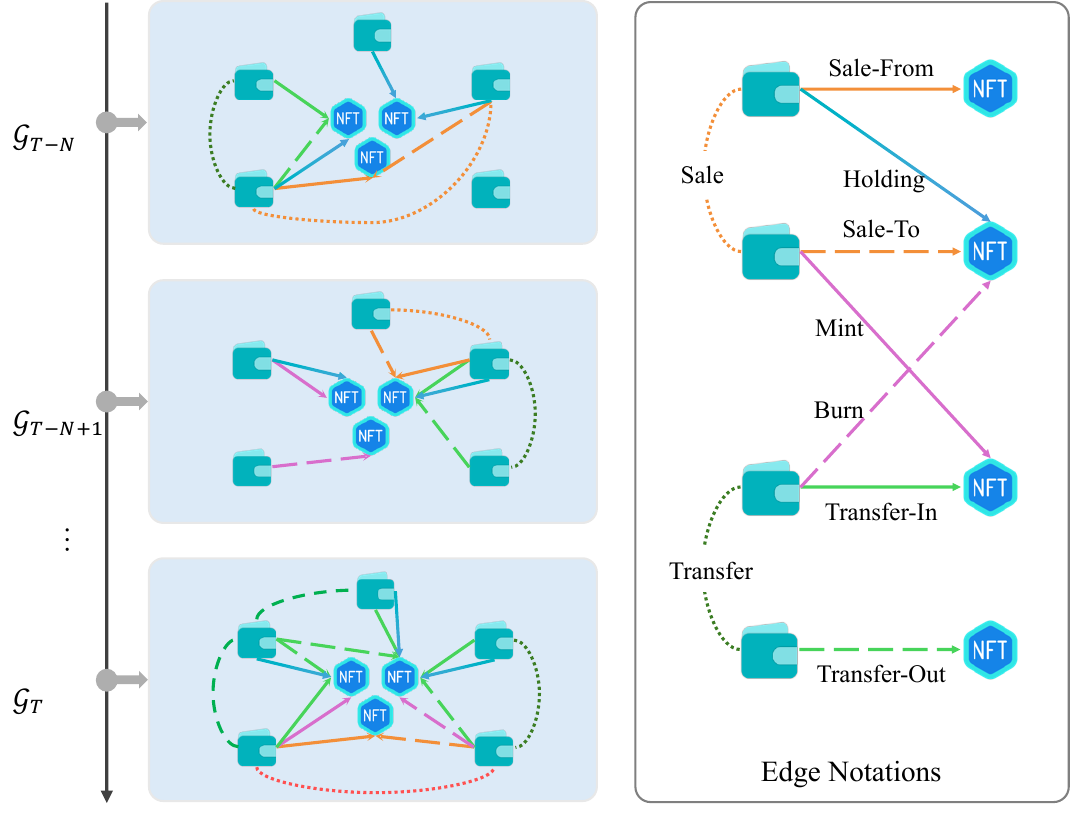}
    \caption{Temporal transaction graph construction.}
    \label{fig:temporal-graph}
\end{figure}

\subsubsection{\text{Data Preprocessing}.}
To improve the quality and usability of NFT-related data, we perform a series of key data preprocessing.

\textit{Sale Noise Removal}.
As is common in financial markets, the NFT marketplace often experiences the presence of noisy transactions. To mitigate this, we first employ the robust strongly connected graph method to identify and isolate wash sales, similar to work \cite{security-ccs-2022-ecosystem}. Then, we apply the Box-Whisker method on sale time series to exclude sales with anomalously high or low prices.

\textit{Collection-level Price Aggregation}. 
In this study, we opt for the daily median price as the preferred collection-level price metric. This choice serves to mitigate the potential impact of anomalous sale prices on the aggregation process, resulting in a more robust modeling of collection-level pricing trends. In cases where there are no recorded sales on certain days, we employ linear interpolation to address the data gap.

\textit{Multimodal Content Embedding}.
NFT consists mainly of visual and textual content. We employ a pretrained vision transformer (ViT) model \cite{dl-lclr-2021-vit} to embed the image of NFTs, resulting in token-level visual embeddings~\cite{chenliyi2022multi}. Regarding textual content, we concatenate various elements, including the collection-level name, description, and all token-level property keys~\cite{generation-mm-2023-profitable-nft}. To derive textual embeddings, we leverage a pre-trained sentence BERT model \cite{dl-emlp-2019-sentence-bert}.

\textit{Token-level Rarity Calculation.} As revealed in studies \cite{valuation-naturesp-2022-heterogeneous,generation-mm-2023-profitable-nft}, the rarity of tokens strongly correlates with price differences among their same collecions. For each collection $c$, we denote its propriety set as $P_c$. We first calculate the rarity score of each property by $r_p = \frac{|c|}{|p|}, p \in P_c$, where $|c|$ denotes the number of tokens in collection $c$ and $|p_c|$ denotes the number of tokens that have the property $p$. Then, the rarity score of a token $i \in I_c$ is defined as $r_i = \sum_{p \in P_c^i} s_p$, where $P_c^i$ is the set of properties of this token.

\subsection{Wallet Profiling Framework}

Wallet profiling aims to create a comprehensive profile of a digital wallet within the NFT ecosystem. This profile encompasses a wide range of attributes and information about wallet holders, contributing to the identification of correlations between their behavior and the NFT price trends they interact with. In general, wallet profiling includes the following four key aspects:
(1) \textit{Investment experience} involves analyzing a wallet's historical transaction frequency, total asset value, and profitability. It provides insights into the wallet holder's level of investment experience, as well as their track record in making successful investment decisions. Wallets with more extensive experience may exhibit more sophisticated trading strategies and a greater impact on NFT prices.
(2) \textit{Content preference} examines the specific styles or types of NFTs that wallet holders are most interested in.
Understanding a wallet holder's content preferences contributes to identifying trends in the popularity of NFT.
(3) \textit{Social connections} explore the latent relationships between wallets within the NFT ecosystem. It provides insights into potential collaborations, shared interests, or contrasting attitudes towards the same NFTs among wallet holders.
(4) \textit{Behavior patterns} focus on the dynamic trading strategies employed by a wallet. This includes assessing how wallet holders react to price fluctuations, whether they engage in speculative trading, or if they adopt a long-term investment approach. Understanding these behavior patterns is critical for evaluating how the actions of wallet holders influence NFT prices over time. Based on this framework, we instantiate a model, namely COMET. 
As shown in the middle of Figure~\ref{fig:enter-label}, COMET contains four parts and we elaborate on each part in the following sections.

\subsubsection{\text{Temporal Transaction Graph Construction}}
The NFT ecosystem is characterized by multifaceted interactions between wallets and collections, indicated by transactions. 
To model the profile of wallets, we construct a temporal transaction graph. Specifically, we segment transactions into daily time windows and build a series of snapshot transaction graphs. We present the transaction graph at $t$ day as $\mathcal{G}_t = (\mathcal{N}_t, \mathcal{R}_t)$, where $\mathcal{N}_t$, and $\mathcal{R}_t$ present the nodes, and edges, respectively. 
\begin{itemize}[leftmargin=*]
\item \textit{$\mathcal{N}_t$}: $\mathcal{G}_t$ consists of two types of nodes, i.e., $\mathcal{W}$ and $\mathcal{C}$, where $\mathcal{W}$ refer to the user's wallets, and $\mathcal{C}$ represent aggregated NFT collections. For $w \in \mathcal{W}$, we consider the daily count of mint, sale, transfer, burn transactions, etc, as their dynamic features $f_w^d$, to model the investment experience of wallets. For $c \in \mathcal{C}$, we consider visual and textual content embeddings as static features $f_c^s$ of each collection, which combines the holding behaviour to acquire the wallets' content preference. Moreover, we allocate each wallet node and collection node a trainable embedding, denoted as $\bar{h}_w$ and $\bar{h}_c$. 
\item \textit{$\mathcal{R}_t$}: There are two main types of edge, i.e., wallet to wallet and wallet to NFT, in transactions, where the former aims to capture the wallets' social connections while the later is to model the behaviour patterns of wallets. A wallet can buy or transfer in an NFT from another wallet, which indicates two relations, i.e., sale $r_s$ and transfer $r_t$, between wallets. A wallet can mint, hold, and burn an NFT, where we denote these three relations as $r_m$, $r_h$, and $r_b$, respectively. Moreover, an NFT is also involved in the sale and transfer transactions between wallets, which presents four relations, i.e., sale-from, sale-to, transfer-in , transfer-out, denoted as $r_{sf}$, $r_{st}$, $r_{ti}$, and $r_{to}$, separately. For the wallet to NFT relations, we merge the NFTs to one node if they belong to the same collection. This can relieve the data sparsity problem at the NFT token level (See Appendix \ref{section:distribution} for detail). 
\end{itemize}

Incorporating the four aspects of wallet profiling when constructing the graph, we can learn comprehensive descriptions of wallets through information extraction from the constructed graph. 
We summarize detailed features of transaction graphs in Appendix \ref{section:features}. By modeling all snapshots within the time window as such transaction graphs, we obtain a temporal transaction graph that evolves at the daily level. We denote it as $\{\mathcal{G}_t \mid t \in [T - H, T] \}$. Our approach provides comprehensive modeling of the dynamic NFT ecosystem.
It enables a holistic understanding of the interplay between collections and wallets, while also implicitly capturing correlations among collections.

\subsubsection{\text{Global Community Detection}}
Within a dynamic NFT ecosystem, the transaction counts of wallets exhibit a characteristic power law distribution, posing a significant challenge of data sparsity (see Appendix \ref{section:distribution} for detail).
Notably, transfer behaviour within the NFT ecosystem tends to exhibit a close relationship between the sender and the receiver wallet \cite{security-ccs-2022-ecosystem}.
For example, there may be different wallets registered by the same user. These wallets tend to exhibit similar market sentiments and behaviour patterns.
To capture these intrinsic connections among wallets, we establish a global wallet transfer graph, based on all historical transfers within the specified time window. In this graph, nodes represent individual wallets, while edges indicate instances of transfer behaviour between these wallets. We measure the strength of the connections between wallets by considering the count of transfers as the weight of each edge.
Subsequently, we employ the Louvain algorithm \cite{ml-jsm-2008-community} to conduct community detection within this graph, resulting in the identification of distinct wallet clusters. These clusters are collectively denoted as $\Phi$, and each wallet $w$ is associated with a specific cluster $\phi \in \Phi$.
This approach enables us to reveal the inherent community structure among wallets, and collaboratively consider market sentiment and dynamic evolution at the community level.

\subsubsection{\text{Transaction Graph Network}}

For snapshot graphs $\{\mathcal{G}_t\}$, we propose a transaction graph network to obtain the embedding of each collection and wallet. There are three components in this network, i.e., node embedding initialization, community-aware information fusion, and multi-behaviour semantic aggregation. 

\textit{Node embedding initialization} provides the initial embeddings for each collection and wallet, which can be presented as follows:
\begin{equation}
    \tilde{h}_{w,0} = \bar{h}_w + \mathrm{MLP}(f_w^d), \quad \forall w \in \mathcal{W},
\end{equation}
\begin{equation}
    h_{c,0} = \bar{h}_c + \mathrm{MLP}(f_c^d) + \mathrm{MLP}(f_c^s), \quad \forall c \in \mathcal{C},
\end{equation}
\noindent where $\bar{h}_w$ and $\bar{h}_c$ are randomly initialized embeddings that will be learned through gradient descending algorithm. 

\textit{Community-aware information fusion} is to address the sparse transaction issue and perceive the community-driven collaboration and evolution.
Specifically, we first derive the embedding $h_\phi$ of each cluster $\phi$ with the mean pooling method,
i.e., $h_\phi = \frac{1}{|\phi|} \sum_{w \in \phi} \tilde{h}_{w,0}, \forall \phi \in \Phi$. Then, we concatenate each wallet node embedding $h_w$ and its cluster embedding $h_\phi$ and encode through MLPs, to integrate the community-level information, and obtain ${h}_{w,0}$. When the context is clear, we unifiedly use $h$ to indicate embeddings of nodes, i.e., wallets and collections, in the following presentations. 

\textit{Multi-behaviour semantic aggregation} is to propagate embeddings along with the constructed multi-type edges
to model the wallet profile and its influence on the price of NFT collections. 
This process consists of multiple graph network layers, and given the $n$-th graph network layer, we consider its input node embedding as $h_{i,n}, \forall i \in \mathcal{N}_t$.
Specifically, for each distinct edge type $r \in \mathcal{R}$, we engage in adaptive semantic propagation to adjacent nodes~\cite{gnn-iclr-2018-gat}. 
\begin{equation}
h_{i,n}^{r} = h_{i,n} + \sum_{j \in \mathcal{N}_i^r} \alpha_{i, j} h_{j,n}, \quad \forall i \in \mathcal{N}_t, \quad \forall r \in \mathcal{R}_t,
\end{equation}
\noindent where $\mathcal{N}_i^r$ denotes the neighbors of node $i$ under the view of edge type $r$. $\alpha_{i, j}$ denotes adaptive weights between node $i$ and $j$. Considering there are features $h_{(i,j),n}^r$ for some edges, we add them with features of node $i$ and $j$ before computing the adaptive weights, i.e., $h_{(i,j),n}^f=h_{i,n}+h_{j,n}+h_{(i,j),n}^r$. For the edges without features, we set their features as zero vectors. Then, $\alpha_{i, j}$ is calculated as follows:
\begin{equation}
    \alpha_{i,j} =
\frac{
\exp\left(\theta_{v,r}^{\top} \sigma (\theta_{w,r} h_{(i,j),n}^f\right)}
{\sum_{k \in \mathcal{N}_i^r}
\exp
\left(\theta_{v,r}^{\top} \sigma (\theta_{w,r} h_{(i, j),n}^f\right)},
\end{equation}
where $\theta_{v,r}^{\top}$ and $\theta_{w,r}$ are trainable parameters, $\top$ denotes the transform operator and $\sigma$ denotes the activation function. 

To aggregate the semantic information from various edge types, we adopt the sum-pool method to integrate each node's representations under different types of edges, calculated as follows:
\begin{equation}
    h_{i,n+1} = h_{i,0} + \sum_{r \in \mathcal{R}} h_{i,n}^r, \quad \forall i \in \mathcal{N}_t.
\end{equation}
Here, we also introduce the residual connection to enhance the original features, while avoiding the over-smoothing issues.

\subsubsection{\text{Temporal Attention Representation}}
After obtaining overall representations of each collection $c$ at different time steps $t$, we use a long short-term memory (LSTM) network to capture temporal patterns among sequential snapshots~\cite{tianfuwang-icc-2021-drl-sfcp,tianfuwang-tsc-2024-hrl-acra}. We can obtain the hidden state $h^t_c$ of each collection $c$ at each timestep $t$:
\begin{equation}
    h_c^t = \mathrm{LSTM} (h_c^{t-1}, x_c^t), \quad \forall {t \in [T - H, T]},
\end{equation}
where $x_c^t$ denotes the hidden state of LSTM at timestep $t$.

Subsequently, we recognize that data from different time steps may have varying degrees of significance in contributing to the overall sequence representation \cite{baseline-ijcai-2017-alstm}. To account for this, we create a representation with temporal awareness by incorporating the attentive weight $\alpha_t$ assigned to each hidden state $h_c^t$, and fusing it with the final hidden state $h_c^T$.
\begin{equation}
    \hat{h}_c = h^{T}_c +  \!\!\!\!\!\! \sum_{t \in [T-H, T]}  \!\!\!\!\!\! \alpha_t h_c^t, \quad \alpha_t = \frac{\exp(\theta_v^{\top}\sigma(\theta_w(h^t_c)))}{\sum_{T \in [T-H, T]} \exp(\theta_v^{\top}\sigma(\theta_w( h_c^t)))}.
\end{equation}

This mechanism ensures that the contributions of different time steps are appropriately weighted and considered in the final representation, acknowledging the varying importance of each snapshot.

\subsection{Hierarchical Prediction and Optimization} 
For two-level tasks, COMET is equipped with two predictors, hierarchically predicting distinct targets and optimizing models.
\subsubsection{\text{Collection-level Price Trend Predictor}}
Given the temporal attention representations of each collection $c$,
we lastly use an MLP-based predictor to generate the predictions of the price movements:
\begin{equation}
    \hat{y_c} = \sigma_c( \mathrm{MLP} (\hat{h}_c))
\end{equation}
where $\sigma_c$ is the $sigmoid$ activation function to produce probability.

\subsubsection{\text{Token-level Sale Price Predictor}} 
For each token $u$, the features associated with each sale $s \in \mathcal{S}_u$ include the sale price, ETH-to-USD exchange rate, and the collection-level price on that specific day, denoted $f_u^s$. Given the heterogeneity of token content, variations in price can exist even within the same collection. To account for these relative price relationships between individual tokens and their respective collections, we draw inspiration from the work~\cite{dl-cikm-2019-bert4rec} and employ a sale sequence encoder based on the Transformer~\cite{dl-nips-2017-transformer} (TFEncoder).
We derive the token sale embedding $h_u^s$ as follows:
\begin{equation}
    h_u^s = \mathrm{TFEncoder}(\{f^s_u \mid s  \in \mathcal{S}_u\}).
\end{equation}
Specially, in the absence of prior sale events for a token, we utilize a zero embedding as its event embedding. 
We also consider the rarity score and total counts of every transaction type as its global features, denoted as $f_u^g$, which is transformed by the MLP, i.e., $h_u^g = \mathrm{MLP}(f_u^g)$.
Finally, we fuse them with the latest hidden state $h_c$ of its collection obtained from the transaction graph network, which contributes to perceiving the future trend of overall collections. We produce the token-level price prediction with the $\mathrm{relu}$ activation function $\sigma_u$ .
\begin{equation}
    \hat{y_u} = \sigma_u( \mathrm{MLP}(h_u^s + h_u^g + \hat{h}_c))
\end{equation}

Considering the dependencies between the two tasks, we optimize COMET with a hierarchical strategy. Concretely, in the training of collection-level task, we minimize the binary cross-entropy (BCE) loss to optimize both the transaction graph network and collection-level predictor. Then, in the training of token-level task, the COMET' parameters that are trained in the collection-level task are frozen. We merely optimize the token-level predictor with the mean squared error (MSE) loss. Moreover, to avoid latent overfitting issues, we include L2 regularization as an additional loss.

\subsection{Practical Deployment}
We have deployed our system in practical environments, which consist of the establishment of data infrastructure and the integration of practical products. See Appendix~\ref{section:deployment} for more details.

\section{Experiments}
We evaluate COMET in tasks at both collection and token levels, comparing it to advanced baselines. We also conduct an ablation study to validate the effectiveness of each component and analyze feature importance. Furthermore, we study the impact of transaction count on performance, placed in Appendix~\ref{section:effectiveness-in-different-collections}.

\subsection{Experiment Setup}
\textbf{Datasets.} We evaluate the performance of COMET using collected NFT data from January 2021 to June 2023. Our evaluation dataset focuses on the top 100 collections, ranked by their sales volume. 
It comprises a total of 1.53 million tokens and 0.74 million wallets, with a total of 7.76 million transactions. 
The distribution of transaction types is as follows: mints (46.02\%), sales (33.30\%), transfers (20.03\%), and burns (0.63\%).
For the collection-level task, we divide the daily price time series of each collection into three subsets: the first 70\% for training, the latest 15\% for testing, and the remaining 15\% for validation. 
Following the setting of work \cite{stocl-kdd-2017-multi-frequency}, we consider prediction steps at intervals of 1, 3, and 5 to assess both short-term and long-term forecasting capabilities. To capture historical trends effectively, we set the size of historical windows to 14. Similarly, the token-level sale dataset is divided according to the same proportions.

\textbf{Baselines.}
For the collection-level classification task, we compare COMET with three groups of baselines: (1) traditional machine learning models, i.e., random forest (RF) \cite{baseline-ml-2001-random-foreset}, support vector machine (SVM) \cite{baseline-ml-1995-svm}, and XGBoost \cite{baseline-kdd-2016-xgboost}; (2) stock movement prediction models, i.e., MLP, LSTM \cite{baseline-ijcnn-2017-lstm}, temporal convelutional network (TCN) \cite{stock-www-2019-tcn} and attentive LSTM (ALSTM) \cite{baseline-ijcai-2017-alstm}; (3) advanced time series forecasting models, i.e., D-Linear \cite{baseline-aaai-2023-dlinear}, and N-BEATS \cite{baseline-iclr-2023-n-beats}, Informer \cite{baseline-aaai-2021-informer} and AutoFormer \cite{baseline-nips-2021-autoformer}.
We adjust the final activation function of these time series forecasting models for our classification task.
For the token-level task, we employ several machine learning regression models, including RF, support vector regression (SVR) \cite{baseline-nips-1995-svr}, and XGBoost. We also utilize the MLP, LSTM, TCN, and ALSTM as baseline sale sequence encoders. In addition to MLP, other methods use ALSTM to encode historical data at the collection level.

\textbf{Metrics.} For the collection-level classification task, we adopt the accuracy (ACC) and Matthews correlation coefficient (MCC) as metrics, widely used in studies on stock trend prediction \cite{stock-www-2022-earning,stock-kdd-2021-accurate}. Besides, we utilize the MSE and mean absolute error (MAE) to measure the prediction performance of token-level task.

\textbf{Implementations.} See Appendix for the details on model implementations, hyperparameter settings, and experiment equipment.

\begin{table}
  \small
  \caption{Overall performance on the collection-level task.}
  \label{tab:collection-level-preformance}
  \begin{tabular}{c|cc|cc|cc}
    \toprule
    \multirow{2}{*}{\textbf{Algorithm}} & \multicolumn{2}{c|}{\textbf{1-step}} & \multicolumn{2}{c|}{\textbf{3-step}} & \multicolumn{2}{c}{\textbf{5-step}} \\
    & ACC $\uparrow$ & MCC $\uparrow$ & ACC $\uparrow$ & MCC $\uparrow$ & ACC $\uparrow$ & MCC $\uparrow$ \\
    \midrule
    RF & 0.5627 & -0.0017 &  0.5476 & -0.003 & 0.5675 & 0.0371\\
    SVM & 0.5678 & 0.0537 & 0.5668 & 0.0197 & 0.5884 & 0.0419 \\
    XGBoost & 0.5640 & 0.0665 & 0.5616 & 0.099 & 0.5755 & 0.1119 \\
    \midrule
    MLP & 0.5513 & 0.0620 & 0.5792 & 0.1365 & 0.5808 & 0.1123 \\
    LSTM & 0.5776 & 0.1372 & 0.5840 & 0.1348 & 0.5843 & 0.1237 \\
    TCN & 0.5781 & 0.1387 & 0.5890 & 0.1386 & 0.5978 & \underline{0.1698} \\
    ALSTM & \underline{0.5893} & \underline{0.1537} & \underline{0.5977} & \underline{0.1681} & 0.5946 & 0.1329 \\
    \midrule
    D-Linear & 0.5663 & 0.0451 & 0.5695 & 0.0427 & 0.5910 & 0.0680 \\
    N-BEATS & 0.5835 & 0.1112 & 0.5902 & 0.1425 & 0.6029 & 0.0262 \\
    Informer & 0.5505 & -0.0173 & 0.5894 & 0.1266 & 0.5841 & 0.0262 \\
    Autoformer & 0.5668 & 0.1063 & 0.5645 & 0.1422 & \underline{0.6079} & 0.1479 \\
    \midrule
    \textbf{COMET} & \textbf{0.6075} & \textbf{0.1861} & \textbf{0.6158} & \textbf{0.2092} & \textbf{0.6214} & \textbf{0.2128} \\
    \bottomrule
  \end{tabular}
\end{table}

\subsection{Collection-level Task Results}
The results for the collection-level task across three different prediction step settings are presented in Table \ref{tab:collection-level-preformance}. Notably, COMET consistently outperforms all baselines, which highlights its superiority. We observe that all methods achieve better performance as the prediction step size increases. This trend indicates the challenge posed by short-term fluctuations compared to predicting long-term trends. Moreover, deep learning-based models (ALSTM, N-BEATS, etc) tend to outperform traditional machine learning models (SVM, XGBoost, etc), emphasizing their capacity for capturing non-linear relationships and their suitability for NFT price prediction tasks.
Time series forecasting models (Informer, Autoformer, etc) exhibit better performance in larger prediction step settings, demonstrating their effectiveness in forecasting long-term trends rather than short-term fluctuations.
Particularly, when compared to suboptimal models, COMET exhibits (3.09\%, 21.08\%), (3.03\%, 24.45\%), and (2.22\%, 25.26\%) improvements in the setting of 1-step, 3-step, and 5-step, respectively, concerning metrics (ACC, MCC).
The greater improvement in MCC over ACC suggests that COMET excels at reducing both false positives and false negatives, highlighting its capability to make well-balanced predictions, especially in such a situation involving class imbalance and asymmetric error costs. By mining the market sentiment behind multiple behaviors, our method more accurately identifies the rise and fall of future price trends. This is particularly valuable in NFT investment scenarios, where incorrect predictions may have significant financial implications.

\subsection{Ablation Study}

\begin{figure}
    \centering
    \includegraphics[width=0.47\textwidth]{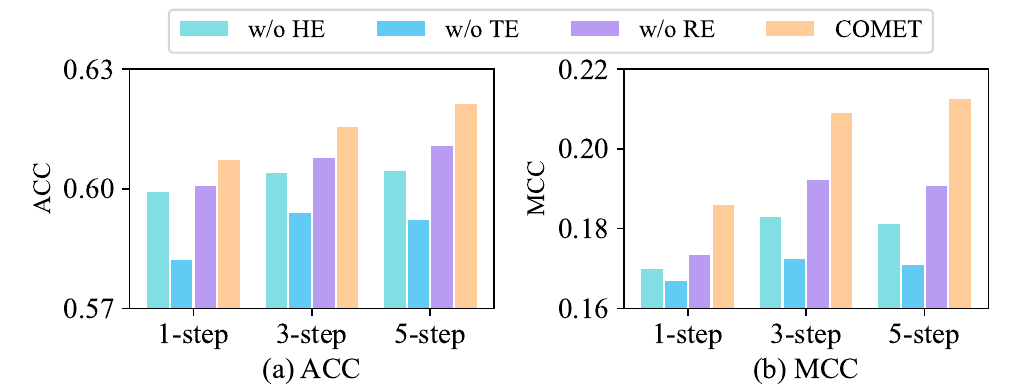}
    \caption{Results of edge ablation.}
    \label{fig:ablation-study-edge}
\end{figure}

\begin{figure}
    \centering
    \includegraphics[width=0.47\textwidth]{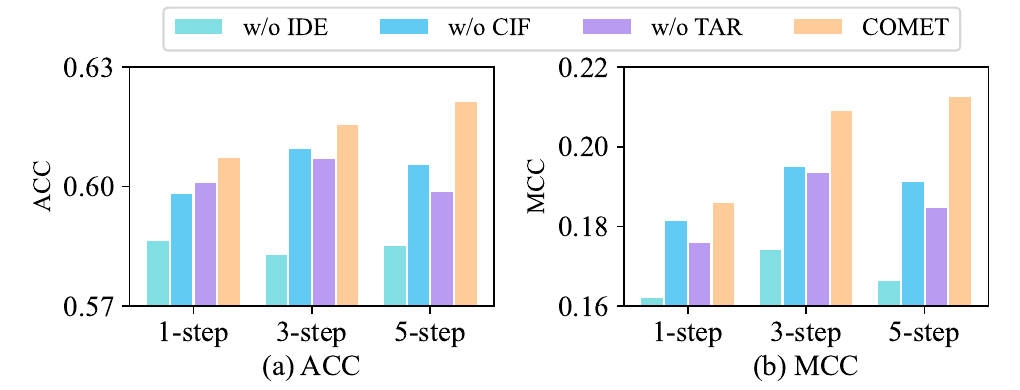}
    \caption{Results of module ablation.}
    \label{fig:ablation-study-module}
\end{figure}

We conduct ablation studies on COMET to evaluate the contribution of each type of edge in the transaction graph construction method and each module within the transaction graph network.

\subsubsection{\text{Edge Ablation}} To examine the contributions of each type of edge, we compare the COMET with its three variations: (1) \textbf{w/o HE} omits all holding edges; (2) \textbf{w/o TE} removes all transaction-related edges; and (3) \textbf{w/o RE} clears all wallet relationship edges. The results of this analysis are illustrated in Figure \ref{fig:ablation-study-edge}, which passes the paired t-test about the significance level. We observe that removing any of these edge types leads to reductions in performance.
Transaction edges, which capture dynamic interactions between wallets and NFT collections, significantly contribute to COMET's effectiveness. They provide valuable insights into the temporal evolution of NFT market and the market sentiment behind multi-behaviour transactions. On the other hand, holding edges capture varying degrees of influence that wallets with different token holdings have on NFT price trends, also enhancing the performance of COMET. Moreover, relationship edges model the relationships among wallets and shed light on the social dynamics, further contributing to COMET's performance. All results demonstrate the effectiveness of our transaction graph transaction method. 

\subsubsection{\text{Module Ablation}} 
We compare COMET against three of its variations to assess the impact of each module:
(1) \textbf{w/o IDE} removes the trainable embeddings of wallets and collections; (2) \textbf{w/o CIF} discards the community-aware information fusion module; and (3) \textbf{w/o TAR} replaces the temporal attention representation module with a single LSTM. Figure \ref{fig:ablation-study-module} reports the results of this study, passing paired t-test. We observe that removing any components causes the performance degradation of COMET, underscoring the indispensable contributions of each component to COMET's overall performance. In particular, among the modules, removing IDE has the most significant decrease in performance. 
This phenomenon can be attributed to the fact that, beyond the transaction history, each wallet and collection possesses inherent characteristics, such as the distinct behavioural patterns exhibited by wallets and the unique attributes associated with collections. These inherent characteristics are also significant drivers of NFT price prediction. IDE module incorporates this personalization into the modeling process, facilitating the learning of these unique characteristics from historical transactions. It enables COMET to achieve superior performance.

\subsection{Feature Importance Analysis}
\begin{figure}
    \centering
    \includegraphics[width=0.45\textwidth]{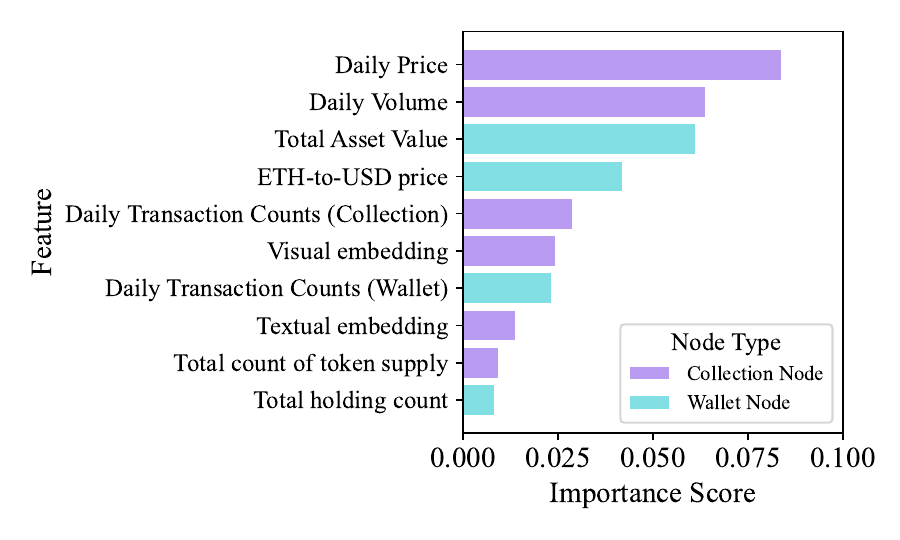}
    \caption{Top 10 feature importances.}
    \label{fig:feature-importance}
\end{figure}

To further assess the impact of utilized features, we employ the permutation importance method \cite{ml-bio-2010-permutation}. Figure \ref{fig:feature-importance} shows the importance score of each feature. 
It is evident from the analysis that historical prices and sale volumes of collections emerge as the two most influential features. These features are crucial because they directly provide historical price trends and trading activities of NFT collections.
Notably, the wallet asset value also ranks high in importance.
This finding suggests that wallets with different asset values have various impacts on NFT price trends.
The behaviours and decisions of wallets with larger asset values may potentially exert a considerable influence on NFT prices.
Interestingly, visual embeddings exhibit higher importance scores compared to textual embeddings, which underscores the higher correlations between visual content and NFT price heterogeneity.

\begin{figure}[!t]
\centering
\includegraphics[width=0.47\textwidth]{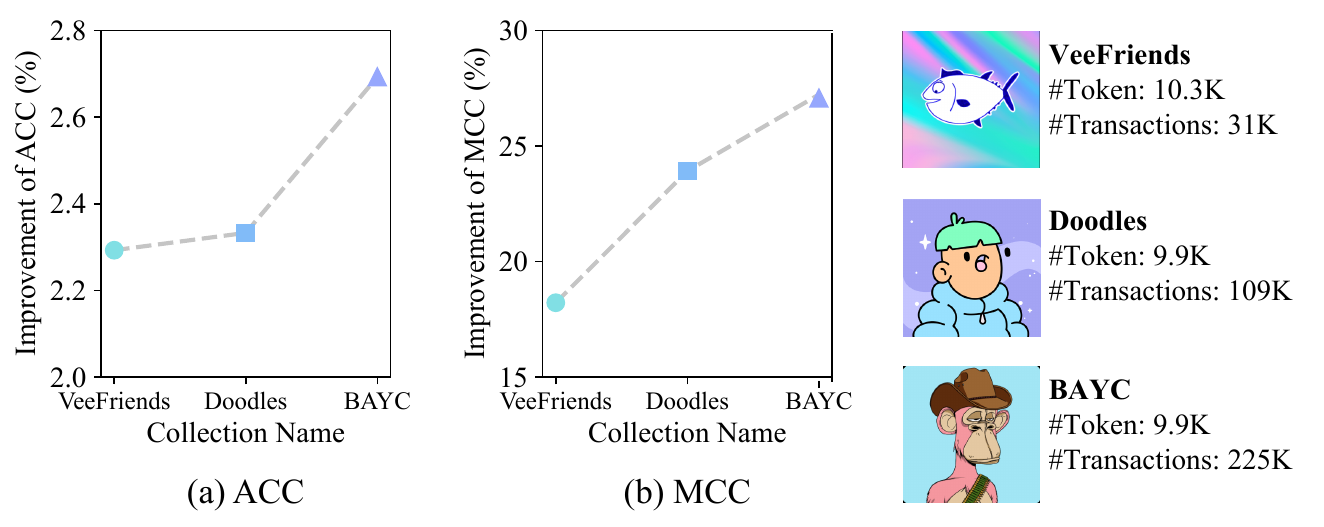}
\caption{Performance improvements of COMET compared to ALSTM cross different collections (5-step).}
\vspace{-0.13cm}
\label{fig:trend}
\end{figure}

\subsection{Effectiveness in Different Collections} \label{section:effectiveness-in-different-collections}

To assess the performance of COMET on collections with varying transaction counts, we compare COMET with ALSTM on three representative collections (BAYC\footnote{Collection address: 0xbc4ca0eda7647a8ab7c2061c2e118a18a936f13d}, Doodles\footnote{Collection address: 0x8a90cab2b38dba80c64b7734e58ee1db38b8992e}, and VeeFrends\footnote{Collection address: 0xa3aee8bce55beea1951ef834b99f3ac60d1abeeb}). 
These collections possess similar token counts but 
differ significantly in transaction counts.
The key distinction between COMET and ALSTM lies in COMET's ability to harness the wallet profiling method to extract NFT price-related insights from transaction behaviors.
In this analysis, we set the prediction step to 5. Figure \ref{fig:trend} shows the improvements achieved by COMET compared to ALSTM across these collections.
We observe that COMET tends to achieve more significant performance improvements in collections with higher transaction counts. This results mainly from the fact that collections with higher transaction counts typically provide richer and more diverse data for COMET. This increased data availability allows COMET to better learn and capture the underlying patterns and dynamics of the NFT market.

\subsection{Token-level Task Results}
Table \ref{tab:token-level-preformance} reports the token-level task results of COMET and all baseline methods. We also compare COMET with two variations: (1) \textbf{w/o CE} eliminates the collection embedding from the transaction graph network of COMET, and instead employs ALSTM to encode the collection-level historical data; and (2) \textbf{w/o TF} replace the Transformer-based sale encoder with an LSTM. 
We observe that the performance differences among all algorithms are relatively small across various prediction settings. This phenomenon can be attributed to the low trading frequency of tokens and the significant time intervals between transactions.
Notably, COMET outperforms all baseline models as well as its variations, which demonstrates the effectiveness of COMET and each component.
Specifically, COMET's transformer-based sale encoder plays a crucial role in capturing the relative relationships between tokens and collections through historical sales sequences. Additionally, the collection embedding from the transaction graph network in COMET provides an overarching perspective at the collection level, integrating historical price trends and market sentiment changes.
This combination of capabilities contributes to COMET's superior performance in token-level price prediction tasks.

\begin{table}
    \small
  \caption{Overall performance on the token-level task.}
  \label{tab:token-level-preformance}
  \begin{tabular}{c|cc|cc|cc}
    \toprule
    \multirow{2}{*}{\textbf{Algorithm}} & \multicolumn{2}{c|}{\textbf{1-step}} & \multicolumn{2}{c|}{\textbf{3-step}} & \multicolumn{2}{c}{\textbf{5-step}} \\
    & MAE $\downarrow$ & MSE $\downarrow$ & MAE $\downarrow$ & MSE $\downarrow$ & MAE $\downarrow$ & MSE $\downarrow$ \\
    \midrule
    RF      & 0.4723 & 2.1812 & 0.4757 & 2.2164 & 0.4743 & 2.1892 \\
    SVR     & 0.4901 & 2.2098 & 0.4849 & 2.1602 & 0.4921 & 2.2178 \\
    XGBoost & 0.4815 & 2.1524 & 0.4950 & 2.2175 & 0.4835 & 2.1604 \\
    \midrule
    MLP     & 0.4181 & 1.9728 & 0.4305 & 2.0321 & 0.4201 & 1.9808 \\
    LSTM    & 0.3689 & 1.775  & \underline{0.3638} & \underline{1.7831} & 0.3679 & 1.7836 \\
    TCN     & 0.3859 & 1.9224 & 0.4003 & 1.9808 & 0.3796 & 1.8904 \\
    ALSTM   & \underline{0.3631} & \underline{1.7523} & 0.3714 & 1.8054 & \underline{0.3601} & \underline{1.7209} \\
    \midrule
    \textbf{COMET}   & \textbf{0.3442} & \textbf{1.5825} & \textbf{0.3379} & \textbf{1.619}  & \textbf{0.3395} & \textbf{1.5643} \\
    \textbf{w/o CE}  & 0.3529 & 1.6163 & 0.3424 & 1.6370 & 0.3479 & 1.6005 \\
    \textbf{w/o TF}  & 0.3576 & 1.6394 & 0.3561 & 1.6793 & 0.3606 & 1.6674 \\
    \bottomrule
  \end{tabular}
\end{table}

\section{Conclusion}
In this paper, we have investigated an emerging application problem, NFT price prediction. To bring clarity to this evolving field, we have provided a practical and hierarchical problem definition encompassing collection-level and token-level tasks, which cater to the diverse practical requirements of investors. 
we also have presented our NFT price prediction system, which includes robust data infrastructure and advanced technical solutions.
Furthermore, we have proposed a novel wallet profiling framework and instantiated the COMET model, enabling us to grasp the influences on NFT prices behind transactions.
Specifically, we have developed a temporal transaction graph construction method that efficiently models the multi-behavior transactions within the NFT ecosystem. Building on this, we have designed a transaction graph network, automatically extracting both temporal dynamics and structural information relevant to NFT prices. Then, we have introduced a hierarchical prediction approach to address the two-level prediction targets.
Extensive experiments on two-level tasks have shown that COMET achieved the best performance, which demonstrates its effectiveness and practical applicability.
Furthermore, we have deployed COMET from the perspectives of both the establishment of the data infrastructure and the integration of practical products.
\begin{acks}
This work was supported by the National Natural Science Foundation of China (Grant No.92370204), National Key R\&D Program of China (Grant No.2023YFF0725001), Guangzhou-HKUST(GZ) Joint Funding Program (Grant No.2023A03J0008), Education Bureau of Guangzhou Municipality, and Nansha Postdoctoral Research Project.
\end{acks}

\clearpage
\bibliographystyle{ACM-Reference-Format}
\balance
\bibliography{main}


\appendix
\section{Appendix}

\begin{figure*}
    \centering
    \subfigure[Data collection pipeline]{\includegraphics[width=0.31\linewidth]{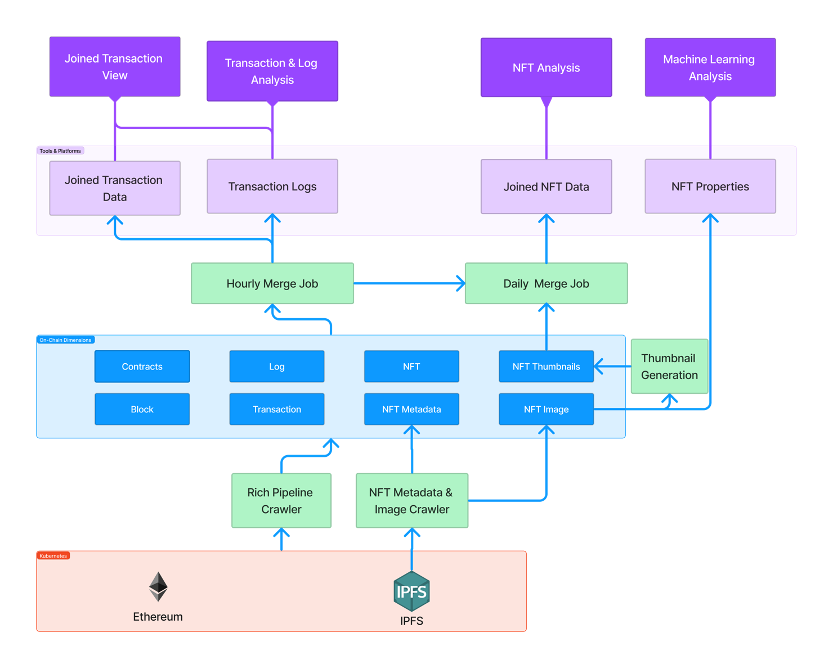}}
    \subfigure[Data visualization panel]{\includegraphics[width=0.61\linewidth]{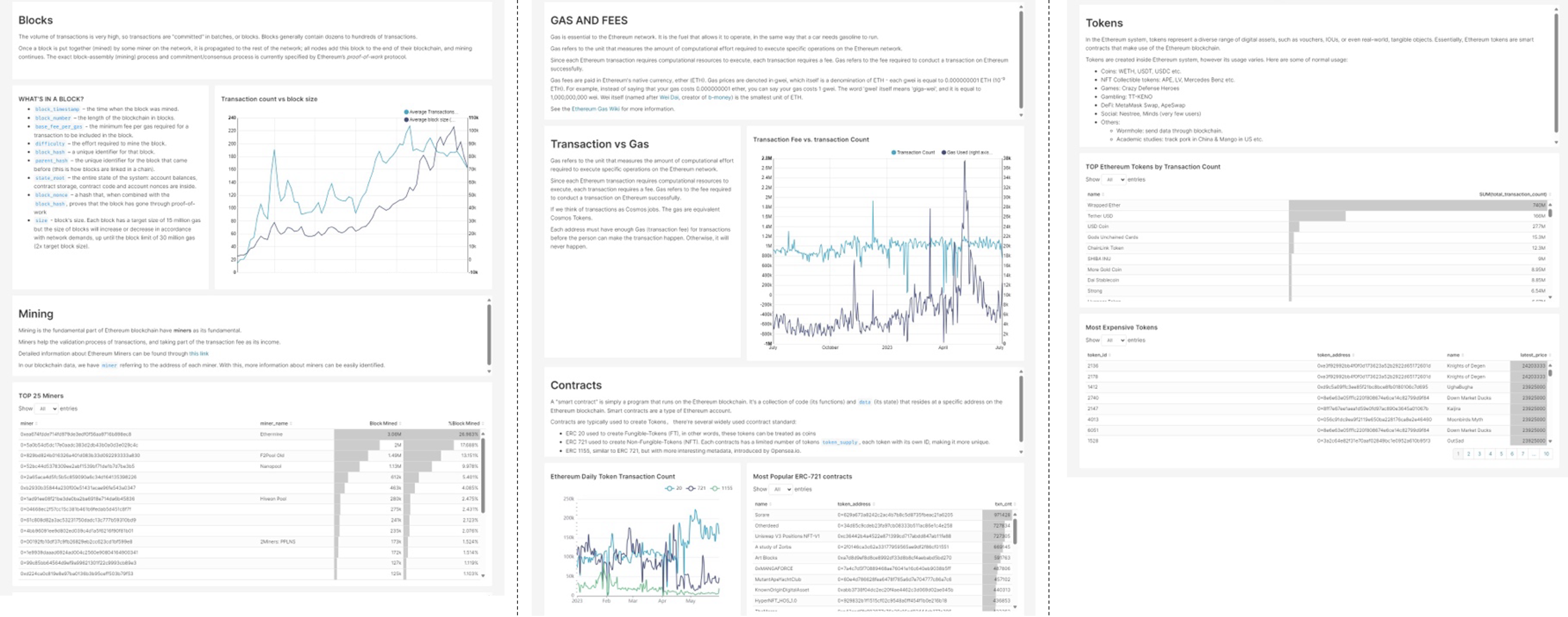}}
    \subfigure[Top collections interface in PowerNFT]{\includegraphics[width=0.46\linewidth]{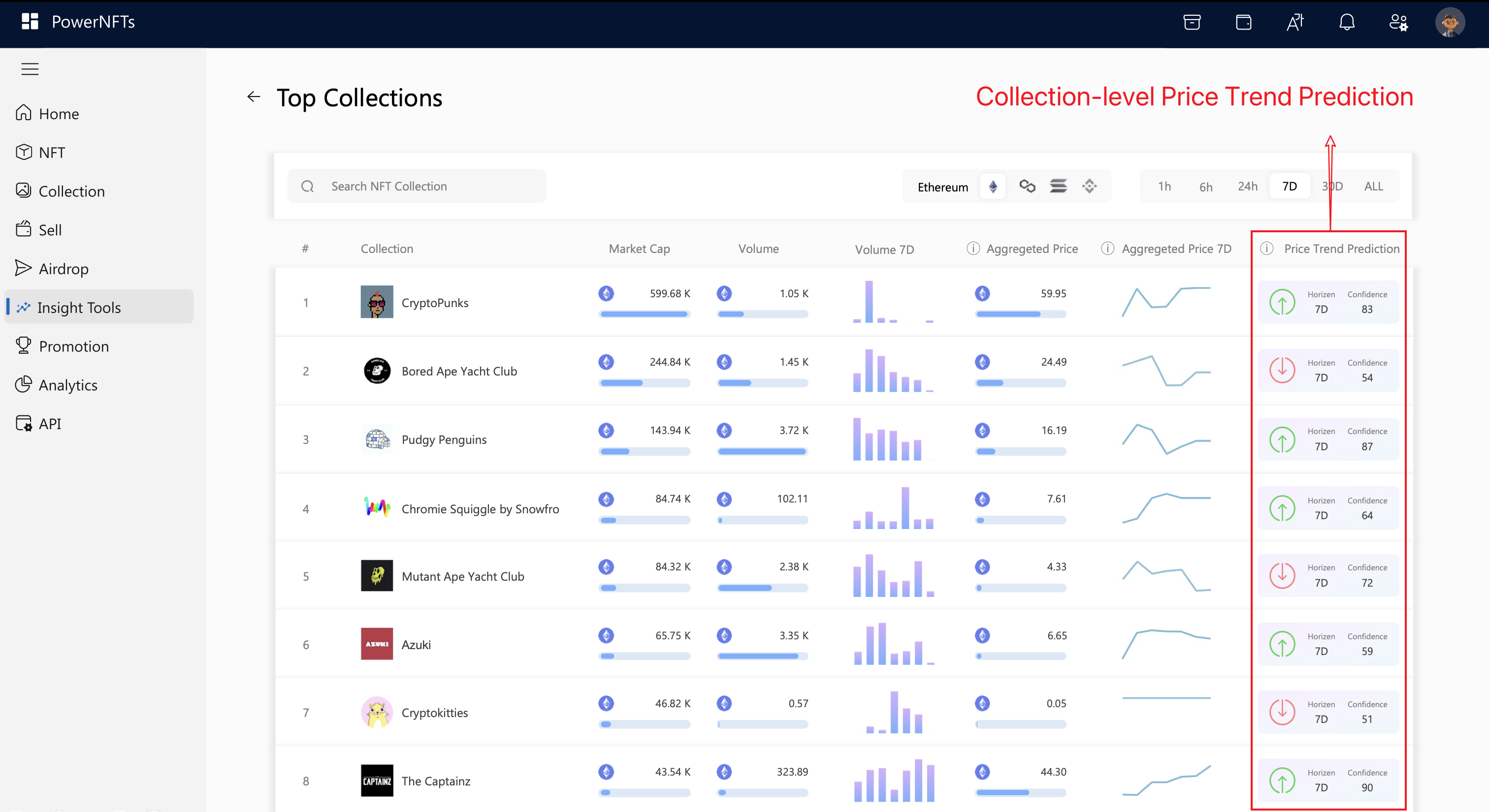}}
    \subfigure[Token details interface in PowerNFT]{\includegraphics[width=0.46\linewidth]{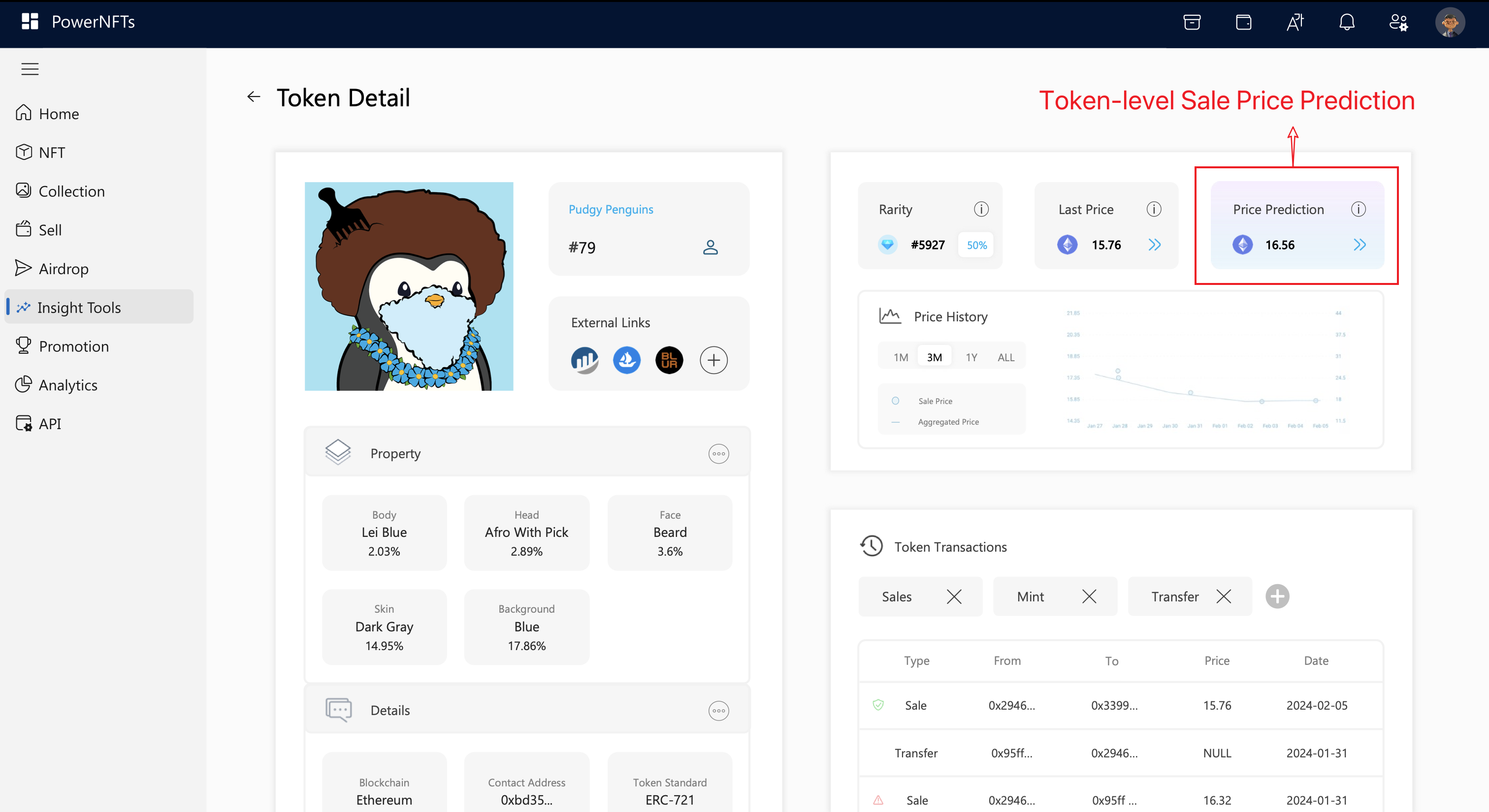}}
    
    \caption{Practical deployment of our system. (a) illustrates the industrial-grade pipeline of data collection. (b) is a dashboard that visualizes the overall blockchain operations. (c) shows the top collections interface in PowerNFT, where price trend prediction is integrated. (d) depicts the token detail interface in PowerNFT, where the predictive sale price is displayed.}
    \label{fig:deployment}
\end{figure*}

\subsection{Deployment Details} \label{section:deployment}
We have deployed our system in Microsoft from the following two aspects: the establishment of data infrastructure and the integration of practical products. 

On the one hand, following the pipeline of data engineering introduced in Section~\ref{data-engineering}, we have constructed an industrial-grade NFT data infrastructure to automate the collection, preprocessing, and visualization of NFT-related data.
As illustrated in Figure~\ref{fig:deployment}(a),
our data collection pipeline utilizes distinct crawlers interfacing with local blockchain nodes and IPFS to acquire both on-chain and off-chain data in real-time, respectively.
Subsequently, we conduct a series of data preprocessing, delivering high-quality data for the NFT-related products.
Additionally, we have developed a comprehensive data visualization panel, showcased in Figure~\ref{fig:deployment}(b).
This interface provides real-time analysis of blockchain metrics related to blocks, contracts, miners, transactions, etc., enabling us to glean valuable insights into overall blockchain operations.
This infrastructure serves as the robust backbone of Microsoft NFT-related products, such as Microsoft crypto wallet and Web3 search, ensuring the efficiency and reliability of data flow.

On the other hand, we have integrated the NFT price prediction model, COMET, into our flagship NFT product, PowerNFT, a platform tailored for building up branded NFT marketplaces. 
Within PowerNFT, COMET functions as an advanced insight tool, offering customers invaluable NFT investment insights and predictive analytics. Figure~\ref{fig:deployment}(c) shows the real-time statistics of top NFT collections in Ethereum, which visualizes the collection information, market cap, sale volume, aggregated price and more. Notably, the "price trend prediction" column adds a layer of foresight, empowering users with future price trend forecasts and confidence levels. Furthermore, as illustrated in Figure~\ref{fig:deployment}(d), our platform goes beyond offering basic token information and historical prices by delivering predictive sale prices, which equip users with actionable insights for informed decision-making. We also introduce visual indicators such as "safe" or "danger" icons in transaction activities, helping users identify potential wash sales.

\subsection{Detailed Features} \label{section:features}

We list the detailed features in Table \ref{tab:features}, which are used in the transaction graph construction method of COMET. We categorize all features into two types: static features are invariant over time, whereas dynamic features may vary from day to day.

\begin{table*}[!tp]
\small
\centering
\caption{Detailed Features of Transaction Graph}
\begin{tabular}{ccll}
\toprule
\textbf{Node / Edge} & \textbf{Feature Type} & \textbf{Feature Name} & \textbf{Description} \\
\midrule
\multirow{7}{*}{Collection Node} 
 & Static & Visual embedding & An embedding representing the visual content associated with the collection. \\
 & Static & Textual embedding & An embedding representing the textual content associated with the collection. \\
 & Static & Total count of token supply & Total number of tokens supplied in the collection. \\
 & Dynamic & Daily price & Daily price of the collection, providing a historical price trend. \\
 & Dynamic & Daily transaction counts & Daily counts of mints, sales, transfers, and burns of the collection. \\
 & Dynamic & ETH-to-USD exchange rate & Daily exchange rate of Ethereum (ETH) to US Dollars (USD). \\
 & Dynamic & Total sale Volume & Total sale volume of the collection over one day. \\
\midrule
\multirow{3}{*}{Wallet Node}
 & Dynamic & Daily transaction counts & Daily counts of mints, sales, transfers, and burns of the wallet. \\
 & Dynamic & Total holding count & Total count of NFT tokens held by the wallet. \\
 & Dynamic & Total asset value & Total value of assets held in the wallet. \\
\midrule
Holding Edge & Dynamic & Owned token count &  Number of NFT tokens in one collection owned by the wallet. \\
\midrule
Sale-from Edge & Dynamic & Sale Price & Price at which an NFT token was sold. \\
Sale-to Edge & Dynamic & Sale Price & Price at which an NFT token was sold. \\
Transfer-in Edge & - & - & - \\
Transfer-out Edge & - & -  & - \\
Mint Edge & - & -  & - \\
Burn Edge & - & - & - \\
\midrule
Sale Edge & Dynamic & Sale Price & Price at which an NFT token was sold. \\
Transfer Edge & - & - & - \\
\bottomrule
\end{tabular}
\label{tab:features}
\end{table*}

\subsection{Dataset Distribution} \label{section:distribution}

\begin{figure}[!ht]
\centering
\includegraphics[width=0.46\textwidth]{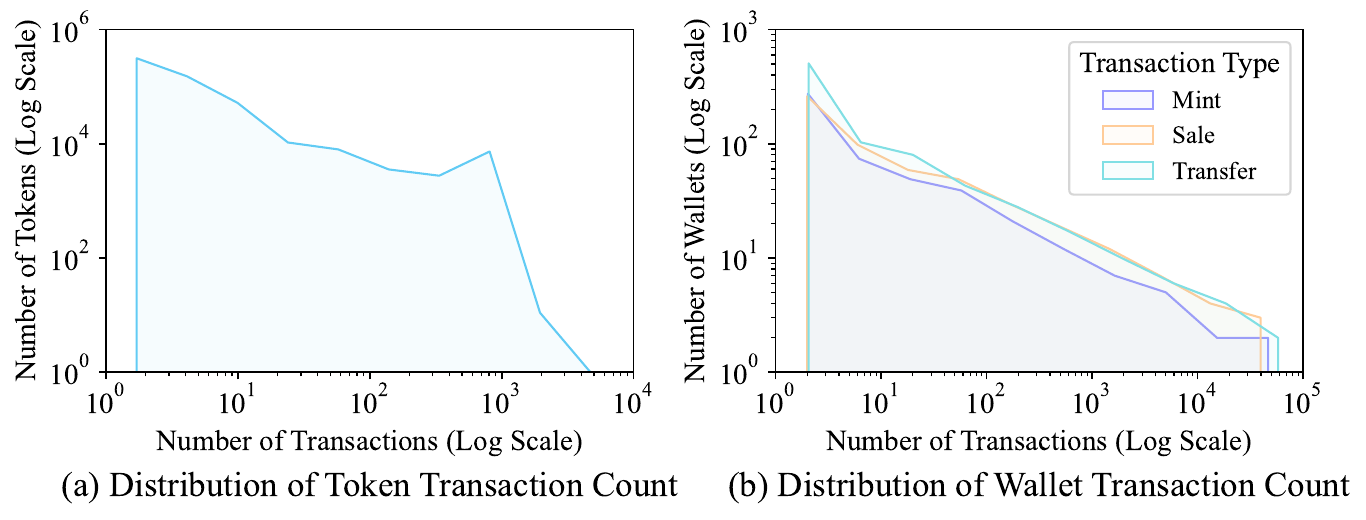}
\caption{Disbutions of token and Wallet transaction counts.}
\label{fig:distribution}
\end{figure}

We present the distribution of the evaluation dataset in Figure \ref{fig:distribution}. As depicted in Figure \ref{fig:distribution}(a), the transaction counts of NFT token exhibits a characteristic of power law distribution. This distribution implies that only a limited number of NFT tokens are actively traded, while the majority remain less frequently transacted. 
Figure \ref{fig:distribution}(b) shows the distributions of transaction counts of wallets, which also follows the power law distribution. This means that only a few wallets actively participate in the NFT marketplace.

\subsection{Implementation Details} \label{section:implementation-details}
We build our neural network model with Pytorch and PyG library. We set the hidden size of neural networks to 64, the number of graph network layers to 2. To alleviate overfitting problems, we set the dropout rate as 0.5 and the weight in L2 loss as 0.0005. Then, we train it using Adam optimizer with a learning rate of 0.001 and a batch size of 64.
Regarding the baselines, for the traditional machine learning model, we implement them with scikit-learn library and use the grid search method to find the optimal candidate hyperparameters. 
For deep learning-based baselines, we adopt the official source code of D-Linear\footnote{https://github.com/cure-lab/LTSF-Linear}, N-BEATS\footnote{https://github.com/philipperemy/n-beats}, Informer\footnote{https://github.com/zhouhaoyi/Informer2020} and AutoFormer\footnote{https://github.com/thuml/autoformer}, and implement others with Pytorch. We integrate the dropout and L2 loss and maintain the most hyperparameters same as our model. Especially, we grid search the learning rate among \{1e-4, 5e-4, 1e-3, 5e-3, 1e-2\} for them and use Adam to optimize. 
All experiments are conducted on a Linux server with a single NVIDIA A100 GPU.

\end{document}